\documentclass[apl, preprint, floatfix,showpacs,amsmath,amssymb]{revtex4}
\usepackage{bbm}
\usepackage{amsfonts}
\usepackage{graphicx}
\usepackage{color}
\usepackage{amsmath}
\usepackage{amssymb}
\usepackage{latexsym}
\usepackage{psfrag}

\begin{document}

\title{Correction factor in non-diffusive Hall magnetometry}

\author{M. Cerchez}\email{mihai.cerchez@uni-duesseldorf.de}\author{T. Heinzel}
\affiliation{Solid State Physics Laboratory (IPkM), Heinrich-Heine-Universit\"at D\"usseldorf, Universit\"atsstr. 1, 40225
D\"usseldorf, Germany}
\date{\today}

\begin{abstract} It is demonstrated how the correction factor $\alpha$ used in
Hall magnetometry of localized magnetic field profiles depends on the sample geometry and on the electron mean free path, in the quasi-ballistic and ballistic regimes, for weak and strong magnetic field regimes. The frequently used approximation of a constant correction factor close to 1 is generally not justified, especially in the case of bipolar magnetic field profiles and may lead to large errors in the determination of the magnitude of the magnetic fields. Rather, $\alpha$ depends in a nontrivial way on the parameters of both the magnetic structure and the Hall cross. 
\end{abstract}
\maketitle

Hall magnetometry has developed into a sensitive and versatile technique to measure
magnetic ($B$) - fields via the Hall voltage they induce in a suitably
designed, cross-shaped probe. While conventional Hall sensors are widely used
in commercial products, \cite{Sze1994} their main area of application in
research is the characterization of magnetic nanopatterns originating from
ferromagnetic
\cite{Johnson1997,Monzon1997,Vancura2000,Kubrak2000,Gallagher2001,Li2002,Reuter2004,Rolff2004,Joo2006,Cerchez2007,Mihailovic2009,Lok1998,Kuhn2000}
or paramagnetic \cite{Mihailovic2007,Kim2008} nanostructures, magnetic domain
walls \cite{Novoselov2003,Christian2006} and superconductors
\cite{Geim1997a,Geim1997b,Pedersen2001}, residing on top of a semiconductor
heterostructure hosting a two-dimensional electron gas (2DEG) about
$100\,\mathrm{nm}$ below the surface. 2DEGs are preferred at liquid helium
temperatures because of their low carrier density and superior noise
characteristics, leading to sensitivities around $10^3$ Bohr magnetons.\cite{Geim1997b} Quantitative magnetometry, however, is
hampered by influences of the magnetic field distribution $B_z(x,y)$, the Hall
cross geometry and the mean free path of the sensing electrons on the measured
Hall resistance $R_H$. This is usually taken into account by a Hall correction
factor $\alpha$, \begin{equation} R_H=\alpha\frac{\langle B_z\rangle}{ne}
\label{eq1} \end{equation} with the average magnetic field in the Hall cross
$\langle B_z\rangle$ and the electron density $n$. Application of Eq. 1 has been limited to the weak magnetic field range, the definition of which, however, has remained somewhat vague. In the diffusive regime, it was shown that $\alpha \leq 1$ \cite{Ibrahim1998,Kim2008}. In the ballistic case, on the other hand, it is established that $\alpha \approx
1$ for $B_z$ profiles well localized inside the Hall cross
\cite{Peeters1998,Lok1998,Kuhn2000} and can become $>1$ for extended $B_z$
fields. \cite{Geim1997b,Christian2006} Many experiments, however, use Hall
sensors in the quasi-ballistic regime i.e., the mean free path $\ell_e$ of the
electrons is larger but still comparable to the size of the Hall cross.
\cite{Johnson1997,Monzon1997,Vancura2000,Kubrak2000,Gallagher2001,Li2002,Reuter2004,Rolff2004,Joo2006,Cerchez2007,Hugger2008,Mihailovic2009}
Here, a model for $\alpha$ is absent. This problem leads to significant errors
in quantitative magnetometry, and the publications either analyze the magnitude
of the Hall resistance qualitatively \cite{Kubrak2000,
Gallagher2001,Li2002,Rolff2004,Reuter2004}, assume implicitly or explicitly
$\alpha =1$
\cite{Johnson1997,Monzon1997,Pedersen2001,Mihailovic2009,Cerchez2007,Hugger2008},
or discuss the data for extremal values of $\alpha$ \cite{Vancura2000}.\\
Studying different magnetic field profiles, we report below that, in the ballistic limit, $\alpha$ can vary by more than a factor of 3 for high magnetic fields, depending on the size of the Hall cross and the magnetic field profile, while for low magnetic fields, $\alpha$ varies only slightly but is not necessarily equal to 1. What is meant by high or low magnetic fields will be naturally classified in terms of the transmission \cite{Kubrak2000PE} of the magnetic field structure. Moreover, the effect of elastic scattering on $\alpha$ is studied in the quasi-ballistic case by numerical simulations based on the semiclassical limit of the Landauer-B\"uttiker approach \cite{Beenakker1989a}.
\begin{figure}[]
\includegraphics[width=85mm]{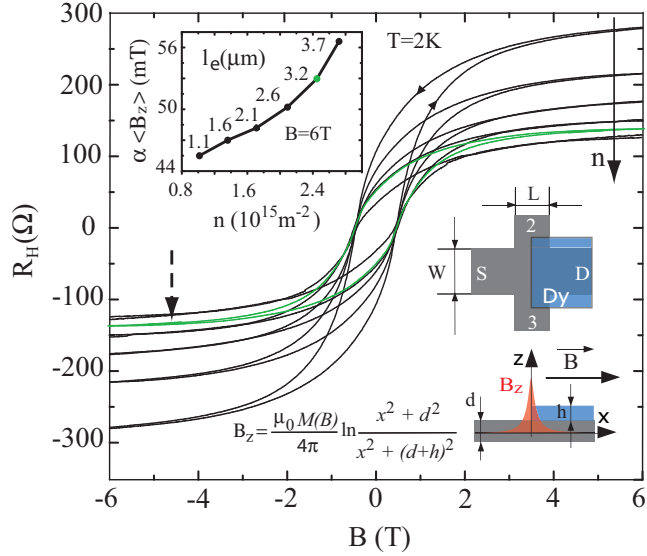}
\caption{Hysteretic Hall resistance as a function of $\vec{B}$ caused by a magnetic
barrier produced under the edge of a Dy film (lower right insets). The
traces are for increasing density as indicated by the solid arrow. Upper left inset: $\alpha \langle B_z\rangle (n)$, indicating that $\alpha$ depends on $n$.}
\label{MCFig1} \end{figure} 
For an experimental illustration of the correction
factor issue, a Hall cross of width $W=10\,\mathrm{\mu m}$ and length
$L=4\,\mathrm{\mu m}$ (inset in Fig. 1) was prepared from a commercially available
$\mathrm{GaAs/Al_{0.3}Ga_{0.7}As}$ heterostructure \cite{IET} with $n=2.45\,\times10^{15}\,\mathrm{m^{-2}}$, a mobility of
$39.8\,\mathrm{m^2/Vs}$ corresponding to an elastic mean free
path of $\ell_e=3.2\,\mathrm{\mu m}$ at a temperature of $2\,\mathrm{K}$. The edge of a ferromagnetic dysprosium film (thickness $h=250\,\mathrm{nm}$) is aligned in transverse $(y-)$ direction at the center of the Hall cross ($x=0$). Magnetizing the Dy film in longitudinal $(x-)$ direction by an external, homogeneous magnetic field $B$ leads to a
characteristic stray field $B_z(x)$. In this paper we will refer to any localized magnetic field profile as \emph{magnetic barrier (MB)}. The structure is covered by a homogeneous top gate used to tune $n$. Measurements are carried out in a liquid helium
cryostat with a variable temperature insert. In Fig. 1, the Hall voltage at a
temperature of $T= 2\,\mathrm{K}$ is shown as a function of $B$. Characteristic
hysteresis loops are observed in the Hall resistance $R_{H}\equiv (V_2-V_3)/I$. Here, $V_j$ denotes the potential of contact $j$ and $I$ the AC current ($100\,\mathrm{nA}$ at a frequency of $17.1\, \mathrm{Hz}$) from the source contact (S) to the grounded drain (D). As $n$ is increased,
$R_{H}(B)$ drops as expected, but (upper left inset) $\alpha$ depends on $n$. As shown below, this is determined by the magnetic landscape and by elastic scattering.
\begin{figure}[]
\includegraphics[width=85mm]{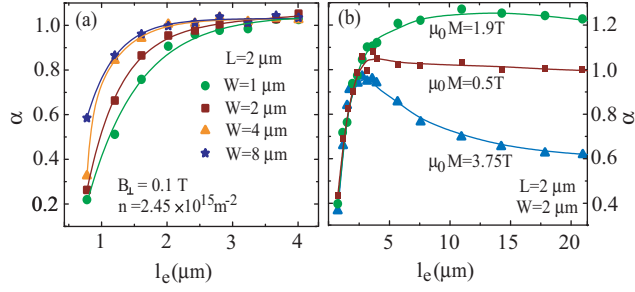} 
\caption{(a) Calculated $\alpha (\ell_e)$ in a homogeneous perpendicular
magnetic field of 0.1 T for different W. (b) $\alpha(\ell_e) $ for the MB introduced in Fig. 1. for 3 MB heights. }
\label{MCFig2} \end{figure}
To demonstrate the effect of scattering in different MB structures, we
calculate $R_H$ using the Landauer-B\"uttiker
formalism \cite{Beenakker1989a} where we have introduced elastic scattering.
We inject from each of the contacts $10^6$ electrons $30\, \mathrm{\mu m}$
away from the Hall cross. The magnetic field region is turned on
adiabatically over a distance of $15\,\mathrm{\mu m}$. Scattering is
introduced by Gaussian distributed small angle scattering, with Poisson
distributed quantum scattering times. Details of the implementation can be found in Ref. \cite{Hugger2007}. This model fails as the diffusive regime is approached, because there is no driving electric field and therefore, a randomized electron
enters any of the Hall probes with equal probability. Hence, the simulated value of $\alpha$ shows an unphysical reduction. This is shown in Fig. 2(a), where $\alpha (\ell_e)$ is calculated for a homogeneous perpendicular magnetic field. Deviations from the expected value $\alpha=1$ set in at $\ell_e\lesssim  L$, thereby defining a lower limit to the range of validity of our model. Fig. 2(b) shows $\alpha(\ell_e)$ for the MB as given by the equation in Fig. 1, centered at $x=0$. We have calculated $\alpha$ for different values of a homogeneous perpendicular field below $0.2\,\mathrm{T}$, with identical results. 
\begin{figure}[] 
\includegraphics[width=85mm]{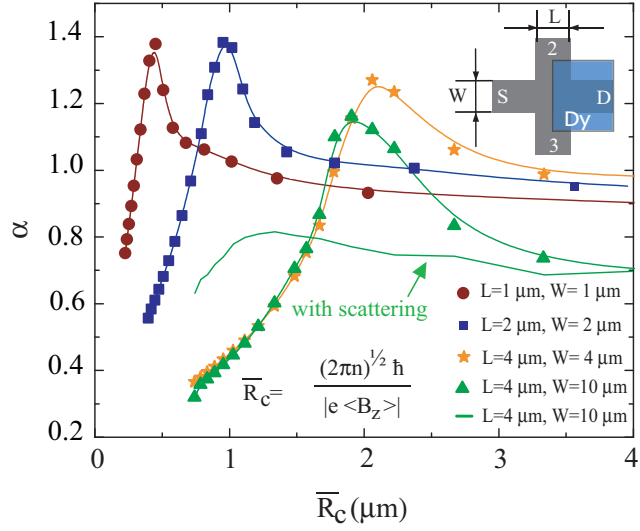}
\caption{$\alpha$ as a function of the cyclotron radius in $\langle B_z\rangle$ for different Hall cross sizes in the ballistic case (symbols with lines), the peaks are at the position where the magnetic barrier closes and the
length of the Hall cross matches the average cyclotron diameter; also shown,
with line, the simulation corresponding to the experimental trace indicated by
the dashed arrow in Fig. 1.}
\label{MCFig3} \end{figure}
\begin{figure}[] 
\includegraphics[width=85mm]{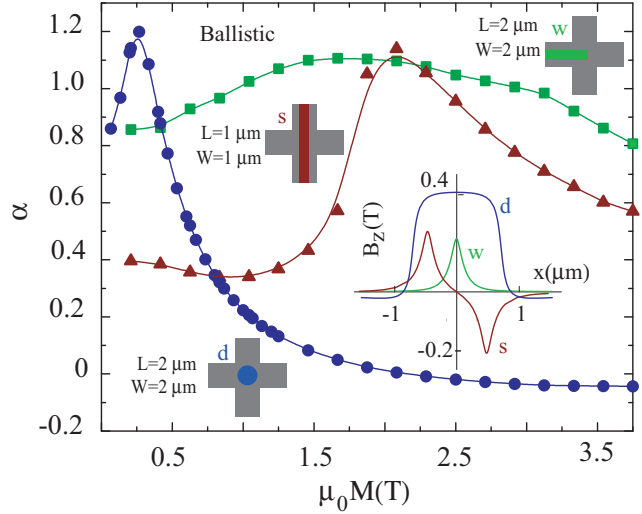}
\caption{The dependence of $\alpha$ on the height of the
magnetic barrier for 3 different magnetic field profiles, magnetic dot d,
stripe s and wire w.}
\label{MCFig4} \end{figure}
For $\ell_e\gg L$, $\alpha$ increases by a
factor of 2 as the MB height is increased, while in the quasi-ballistic regime ($\ell_e\gtrsim L$), $\alpha$ depends strongly on $\ell_e$ while its sensitivity on the MB height drops. To look into detail at the effect of the MB height on $\alpha$, we
have represented $\alpha$ as a function of the cyclotron radius $\bar{R_c}$ the electrons would have in a homogeneous magnetic field equal to $\langle B_z\rangle$, see Fig. 3. For high values of $\bar{R_c}$, $\alpha$ has values between 0.7 and 1 depending on the size of the Hall cross. As $\bar{R_c}$ is reduced, more of the electrons ejected from the source contact are deflected by the MB into contact 2 until the MB closes \cite{Kubrak2000PE} when $e/(\hbar \sqrt{2\pi n}) \cdot \int B_z(x)dx \ge 2$. At this point, which for a MB situated inside the Hall cross corresponds to $2\bar{R_c}=L$, $\alpha$ has a maximum as contact 2 collects most of the electrons. For even lower $\bar{R_c}$, more electrons will be rejected by the MB, such that increase of $R_H$ with the MB height slows down and hence $\alpha$ drops. Taking this behavior into account we can now link the low field regime to the range of slow variation for $\alpha$, when the barrier is mostly transparent, which translates, for a MB inside the Hall cross, in $\bar{R_c}\ge L$ and, in contrast, high magnetic fields start at the point where the MB starts becoming more opaque at $\bar{R_c}< L$.  As scattering is introduced, this strong dependence of
$\alpha$ on $\bar{R_c}$ smears out due to two mechanisms: first, many electrons coming from source which were rejected in the ballistic case by the magnetic barrier, can now end up in contact 2 assisted by a scattering event, with higher probability than in contact 3 due to the curved trajectory in the MB field towards contact 2 . Second, a reduction of the MB height supports electrons to reach the drain helped by a scattering event, with the effect of lowering $\alpha$. The line without symbols in Fig. 3 shows the calculated $\alpha (\bar{R_c}) $ for the experimental parameters present in the trace indicated by the dashed arrow in Fig. 1. Here, $\alpha$ assumes values below 1 in the whole range and varies by $\approx 15\,\%$.\\
We proceed by numerical studies of $\alpha$ for three additional, ballistic geometries reported in the literature.\cite{Novoselov2002,Kubrak2000,Reuter2004} Let us first consider a
bipolar MB structure as produced by a magnetic stripe aligned along the $y$ - direction direction and magnetized to the magnetization $\mu_0M$ in $x$-direction, \cite{Kubrak2000} marked with \textit{s} in Fig. 4. Here, $B_z(x)$ is antisymmetric about $x=0$ and hence, $\langle B_z\rangle=0$, emphasizing the limited usefulness of Eq. 1. We have therefore represented $\alpha$ for $B_z (x)$ averaged only over the interval of positive polarity ($-L/2\leq x\leq 0$). It has a maximum at $\mu_0M=1.87 \,\mathrm{T}$ where the magnetic barrier becomes opaque for the transmission from source into drain. A second example is a
magnetic dot marked with \textit{d}, magnetized in z-direction, centered in the Hall cross and with sample parameters similar to those given in Ref. \cite{Novoselov2002} namely a Dy cylinder of $1.5\,\mathrm{\mu m}$ height and $1.5\,\mathrm{\mu m}$ diameter. In this MB, $B_z$ directly below the dot is of opposite sign to that one further away from the dot in the plane of the 2DEG.
Here, $\alpha$ shows a maximum at a magnetization very close to the point where the MB closes in $x)$ direction along the line $y=0$. Note that $\alpha$ can even become negative for large magnetizations, which is due to the increasing 
negative magnetic field outside but near the Hall cross. Finally we studied a magnetic wire of $0.5\,\mathrm{\mu m}$ width, aligned in the center of the Hall
cross parallel to the transport direction \cite{Reuter2004}, such that a unipolar magnetic dot is formed. Here, $\alpha$ depends weakly on the magnetization assuming values between 0.8 and 1.1 with a broad maximum.  \\ 
To conclude, we have demonstrated theoretically and experimentally that the correction factor in Hall magnetometry is not constant but rather depends on the magnetic field shape and amplitude, on the geometry of the Hall cross and on elastic scattering. For bipolar magnetic field profiles, the generally used Eq. 1 becomes inapplicable even when parametric variations of $\alpha$ are allowed. It has emerged that calculation of the response of the Hall sensor within the presented model can greatly increase the accuracy of quantitative Hall magnetometry in the non-diffusive regime, opening the door for investigating further factors not considered here, like, for example, deviations from the assumed analytic magnetic field profiles.

\newpage

\end{document}